\documentclass[12pt]{article}
\usepackage{graphicx,amsmath,amssymb,cite}
\usepackage{epsfig,epsf}
\usepackage{graphicx}
\usepackage{epstopdf}
\usepackage{cite}
\usepackage[utf8x]{inputenc}

\newcommand{\beq}{\begin{equation}}
\newcommand{\eeq}{\end{equation}}

\numberwithin{equation}{section}

\usepackage{textcomp}

\begin{document}

\begin{center}

{\Large{\bf Dipole model analysis of high precision HERA data  }}

\vspace*{1 cm}

{\large A. Łuszczak$^1$, H. Kowalski~$^2$ }\\ [0.5cm]

{\it $21$ T. Ko\'sciuszko Cracow University of Technology}\\[0.1cm] 
{\it $21$ Deutsches Elektronen-Synchrotron DESY, D-22607 Hamburg, Germany}\\[0.1cm] 
 \end{center}

\vspace*{3 cm}

\begin{abstract}
 We analyse, within a dipole model, the inclusive DIS  cross section data, obtained from the combination of the H1 and ZEUS  HERA measurements. We show that these high precision data are very well described within the dipole model framework, which is  complemented with a valence quark structure functions. We discuss the properties of the gluon density obtained in this way.  \end{abstract}

%\maketitle

%%%%%%%%%%%%%%%%%%%%%%%%%%%%%%%%%%%%%%%%%%%%
%% MAINMATTER
%%%%%%%%%%%%%%%%%%%%%%%%%%%%%%%%%%%%%%%%%%%%
\section{Introduction}

Many investigations have shown that  HERA inclusive and diffractive DIS cross sections are very well described by the dipole models~\cite{BGK,GBW, Iancu:2003ge}.  Interest in the dipole description emerge from the fact that dipole picture provides a natural description of QCD reaction in the low-$x$ region. Due to the optical theorem,  dipole models allow a simultaneous description of many different physics reactions, like inclusive DIS processes, inclusive diffractive processes, exclusive $J/\psi$, $\rho, \phi$ production, diffractive jet production, or diffractive and non-diffractive charm production. In the dipole picture, all these processes are determined by the same, universal, gluon density~\cite{MSM,KT,KMW}.

 In the era of  the LHC, the precise knowledge  of gluon density is very important because the QCD-evolved gluon density determines the cross sections of  most relevant   physics processes, e.g. Higgs production. Any significant deviation of the predicted cross section from their Standard Model value could be a sign of new physics. 

  The validity of the dipole approach was experimentally established, a decade ago, by a  comparison of the dipole predictions with HERA $F_2$ and diffractive data in the low $x$ region~\cite{GBW}, \cite{BGK}.  In the meantime, the precision of data obtained from HERA experiments increased substantially. The H1 and ZEUS experiments have combined their inclusive DIS cross sections which, due to a substantial reduction of systematic measurements errors, led to an increase of precision by about a factor two~\cite{H1ZEUS}. In the same way the quality of the inclusive charm data was substantially improved~\cite{H1ZEUS-charm}. Finally, recently, the exclusive $J/\psi$ production was much more precisely measured~\cite{H1-J/psi} . All these reaction were used in the past to establish the dipole approach. It is therefore interesting to re-evaluate these reactions because the dipole picture provides a somewhat different approach to the gluon density than the usual pdf approach.  In the usual pdf approach the gluon density contributes to  $F_2$ mainly through the evolution of the see quarks, the direct gluon contribution is only of the order of a few percent. On the other hand, in the  dipole models the  gluon density is directly connected to the see quarks. In the pdf scheme the evolution is evaluated in the  collinear approximation whereas  the dipole approach uses the $k_T$ factorization. 
  
  The direct connection between the dipole production and gluon density  is particularly clearly seen in the exclusive $J/\psi$ production, which was therefore proposed as a testing ground of the properties of the gluon density~\cite{MarRysTeu0}.   Presently, the exclusive $J/\psi$ production is precisely measured in heavy ion collisions at RHIC and LHC. These measurements combined with their dipole analysis can become a new source of information about the gluonic structures of nuclei~\cite{CK,KLV}   

 Another important application of the dipole description is the investigation of the gluonic high density states. These can be characterized by the degree by which a dipole is absorbed or multiply scattered in such states. The states with the highest gluon densities are produced today in the high energy  heavy ion scattering at RHIC and LHC.  This is now a very lively field of saturation investigation~\cite{YKov,RV}.
 % However, in this paper we will only partly address the saturation issues since we concentrate here on new HERA data, which are outside of the  region  mostly relevant  to saturation . 

The aim of this paper is to investigate the additional information which is contained in the new, combined HERA data.  The most precise data where obtained in the region of higher Q$^2$'s (Q$^2$ from 3.5 to O(10000) GeV$^2$),  where the DGLAP evolution is known to describe  data very well.  
Therefore, as discussed below,  in this investigation we use the so called BGK dipole model, because it uses the DGLAP evolution scheme.  

This paper concentrates first on  the  {\it inclusive} DIS measurements in the low $x$ region.   
%In this region  the dipole and  the usual pdf approach are equivalent, as  proven in  perturbative QCD (cite). 
Here, the contribution of the valence quarks is small, below 7\%, and has therefore been neglected until now. However, the combined H1 and ZEUS HERA data achieve however a precision of about 2\%, so the contribution of the valence quarks can no longer be neglected. The present paper addresses  the question to what extent the contribution of the valence quark and the dipoles are compatible with each others. To do so we use the HARAfitter framework~\cite{HERAfitter} which allows to treat consistently   QCD evolution together with the  valence quark and dipoles contributions.

 The paper is organized as follows: in Section 2 we  recall the main properties of the dipole approach and review various models in order to  motivate our choice. In Section 3 we discuss the results of fits and in Section 4 we compare the fits with data. Section 5 contains the summary.

%The main motivation for this study is provided by the observation that in the usual pdf approach  the gluon density is not very well defined since the $F_2$   measurements are not very sensitive to its contribution. The gluon density contributes to  $F_2$ mainly through the evolution of the see quarks, the direct contribution is only of the order of a few percent.
%This is different than in case of $F_L$, which in the collinear factorization approach, is mainly determined by the  gluon density.
%  In  case of $F_2$ the uncertainties due to higher order QCD effects are of the order of $10\%$, whereas for $F_L$ they are much larger, of the  order of $100\%$.  This indicates that gluon density is not very well defined within the usual pdf scheme.   On the other hand, in the  dipole models the  gluon density is directly connected to the see quarks and therefore also to $F_2$.  
%This interpretation is confirmed by the successful predictions for  $F_L$ obtained from the $F_2$ measurements analyzed within the dipole models. 

%The different role of the gluon density in the two approaches 
 %could show a way to improve the understanding of the physical properties of the gluon density.  

\section{Dipole models}
In the dipole picture the deep inelastic scattering is viewed as a two stage process; first the virtual photon fluctuates into a dipole, which consists of  a quark-antiquark pair (or a $q\bar q g$ or $q\bar q gg$ ... system) and in the second stage the dipole interacts with the proton~\cite{NNZ:91}
,\cite{
Nemchik1996,Gotsman1995,Dosch1996,CS,Forshaw2003,
Frankfurt2005,KLMV}. Dipole denotes a quasi-stable quantum mechanical state, which has a very long life time ($\approx 1/m_p x\;$) and a size $r$, which remains  unchanged during scattering. The wave function $\Psi$ determines the probability to find a dipole of size $r$ within a photon. This probability depends on the value of external $Q^2$ and the fraction of the photon momentum carried by the quarks forming the dipole, $z$. Neglecting the $z$ dependence, in a very rough approximathion, $Q^2 \sim 1/r^2$.

  The scattering amplitude is a product of the virtual photon wave function, $\Psi$, with the dipole cross section, $\sigma_{\text{dip}}$, which determines a probability of the dipole-proton scattering. 
Thus, within the dipole formulation of the $\gamma^* p$ scattering     
\begin{equation}
\label{edipole}
   \sigma_{T,L}^{\gamma^* p}(x,Q^{2}) = \int dr^2 \int dz \Psi^*_{T,L} (Q,r,z) \sigma_{\text{dip}}(x,r) \Psi_{T,L}(Q,r,z),
\end{equation}
where $T,L$ denotes the virtual photon polarization and $\sigma_{T,L}^{\gamma^* p}$ the total inclusive DIS cross section.

Several dipole models have been developed to test various aspects of the data.  They vary due to different assumption made about the physical behavior of  dipole cross sections.  In the following we will shortly review them and motivate our the choice of the model used for the present investigation.   

%  In the HERAFitter  three representative models  are implemented.
%\begin{itemize}
%\item
%the original (GBW)~\cite{Golec-Biernat:1998js} dipole saturation model,
%\item
%  the colour glass condensate approach
%to the high parton density regime (IIM)~\cite{Iancu:2003ge},
%\item
 % a modified GBW model which takes into account the effects of  
%DGLAP evolution (BGK)~\cite{Bartels:2002cj}.
%\end{itemize}
%%%%
\subsection{GBW model}

The dipole model became an important tool in investigations of deep-inelastic scattering due to the initial observation of Golec-Biernat and W\" uesthoff (GBW) \cite{GBW}, that a simple ansatz for the dipole cross section was able to describe simultaneously the total inclusive and diffractive cross sections.

In the GBW model the dipole-proton cross section $\sigma_{\text{dip}}$ is given by
\begin{equation}
\label{eGBW}
   \sigma_{\text{dip}}(x,r^{2}) = \sigma_{0} \left(1 - \exp \left[-\frac{r^{2}}{4R_{0}^{2}(x)} \right]\right),
\end{equation}
where $r$ corresponds to the transverse separation between the quark and the antiquark, and $R_{0}^{2}$ is 
an $x$ dependent scale parameter which has a meaning of saturation radius,  $R_{0}^{2}(x)=\left(x/x_{0}\right)^{\lambda_{GBW}}$.
The free fitted parameters are: the cross-section normalisation, $\sigma_{0}$, as well as $x_{0}$ and $\lambda_{GBW}$.  In this model saturation is taken into account in the eikonal approximation and the saturation radius is intimately related to the gluon density in the transverse plane, see below. The exponent $\lambda_{GBW} $ determines  the growth of the total and diffractive cross section with decreasing $x$. For dipole sizes which are large in comparison to the saturation radius, $R_0$, the dipole cross section saturates by approaching a constant value $\sigma_0$, i.e. saturation damps the growth of the gluon density at low $x$. 

The GBW model provided a good description of data from medium $Q^2$ values ($\approx 30$ GeV$^2$) down to low $Q^2$ ($\approx 0.1$) GeV$^2$).
Despite its success and its appealing  simplicity  the model has some  shortcomings; in particular it describes the QCD evolution by a simple $x$ dependence,  $ \sim (1/x)^\lambda_{BGW}$, i.e the $Q^2$ dependence of the cross section evolution is solely induced by the saturation effects. Therefore, it does not  match with DGLAP QCD evolution, which is known to describe data very well from $Q^2 \approx 4$ GeV$^2$ to very large $Q^2 \approx 10000$ GeV$^2$.

\subsection{BGK model}
The evolution ansatz of the GBW model was improved in the model proposed by Bartels, Golec-Biernat and Kowalski, (BGK)~\cite{BGK},  by taking into account the  DGLAP evolution of the gluon density in an explicit way. The model preserves the GBW eikonal approximation to saturation and thus the dipole cross section is given by
\begin{equation}
\label{eBGK}
   \sigma_{\text{dip}}(x,r^{2}) = \sigma_{0} \left(1 - \exp \left[-\frac{\pi^{2} r^{2} \alpha_{s}(\mu^{2}) xg(x,\mu^{2})}{3 \sigma_{0}} \right]\right).
\end{equation}
The evolution scale $\mu^{2}$ is connected to the size of the dipole by $\mu^{2} = C/r^{2}+\mu^{2}_{0}$. This assumption allows to treat  consistently the contributions of large dipoles  without making the strong coupling constant, $\alpha_s(\mu^2$),  un-physically large.
  
 The gluon density, which  is parametrized  at the starting scale $\mu_{0}^{2}$, 
is evolved to larger scales, $\mu^2$, using LO or NLO DGLAP evolution.
 We consider here three forms of  the gluon density:
\begin{itemize}
\item 
the {\it soft} ansatz, as used in the original BGK model 
\begin{equation}
   xg(x,\mu^{2}_{0}) = A_{g} x^{-\lambda_{g}}(1-x)^{C_{g}},
\label{gden-soft}
\end{equation}
\item
the {\it soft + hard} ansatz
\begin{equation}
   xg(x,\mu^{2}_{0}) = A_{g} x^{-\lambda_{g}}(1-x)^{C_{g}}(1+D_g x +E_gx^2),
      \label{gden-softhard}
\end{equation}
\item
the {\it soft + negative} gluon 
\begin{equation}
   xg(x,\mu^{2}_{0}) = A_{g} x^{-\lambda_{g}}(1-x)^{C_{g}} -A'_{g} x^{-\lambda'_{g}}(1-x)^{C'_{g}},
      \label{gden-softneg}
\end{equation}
\end{itemize}  
The free parameters for this model are $\sigma_{0}$, $\mu^{2}_{0}$ and the  parameters for gluon $A_{g}$, $\lambda_{g}$, $C_{g}$ or additionally $D_g, E_g,$ or $ A'_g, \lambda'_g,C'_g.$ 
 Their values are obtained by a fit to the data. The fit results were found to be independent on the parameter  $C$, which was therefore fixed as  $C=4$ GeV$^2$, in agreement with the original BGK fits.

\subsection{IIM model}
Although we do not use the IMM   (Iancu, Itacura and Mounier) model in this paper we mention  it here because it may take better into account the saturation effects than it is the case in the BGK or GBW models. The last models use for satuartion the eikonal approximation,  whereas the IIM model uses a simplified version of the  
Balitsky-Kovchegov equation~\cite{Balitsky:1995ub}. The explicit formula for $\sigma_{\text{dip}}$ 
can be found in~\cite{Iancu:2003ge}.  We do not use this model  because we concentrate here on the higher $Q^2$ data, which precise description requires an equally precise transition to the DGLAP regime. 
%%%%

\subsection{Dipole model with valence quarks}
The dipole models are valid in the low-$x$ region where the valence quark contribution is small.  Therefore, this contribution  was usually neglected which  was justified as long as the experimental errors were relatively large.  Theoretically, it is very difficult to treat valence quarks inside the dipole framework because, until now,  the dipole amplitudes are not well defined in the region of high $x$. The problem may be solved, in future, by the analytic continuation of the dipole (or BFKL) amplitudes from the low $x$ to the  high $x$ region~\cite{KLRn}. However,  for the purpose of this paper, we propose to take an heuristic approach and just to  add the valence quark contribution from the standard pdf's fits to the dipole predictions. In this approach the dipole contribution plays a role of the see quarks in the standard pdf's. This procedure is justified by the fact that the see quark contribution disappears at larger $x$. The HERAfitter project is well suited  for this purpose since the dipole model and the valence quarks contributions are a part of the same framework. 
 
\section{Results from fits}
\begin{table}[h]
\begin{center}
\begin{tabular}{|c||c||c||c|c||c|c|c||c|c|c||c|} 
\hline 
No& 
$Q_0^2$ &
$\sigma_0$ & $A_g$ & $\lambda_g$ & $C_g$ & $C$ &  $Np$& $\chi^2$& $\chi^2/Np$\\
\hline
% h09a & 
1 &
$1.1$ &  143.14  & 1.605 & -0.056 &  5.884 & 4.0 &  201 & 198.17 & 0.986 \\
\hline
3 & 
$1.3$ & 123.18 & 1.589 & -0.094& 6.937 & 4.0 & 201 & 200.70 &0.998 \\
\hline
5 &
$1.5$ & 112.44& 1.685 &-0.109& 8.124& 4.0 &  201&  202.26 &  1.006 \\
\hline
7 &
$1.7$ &97.91& 1.603 &-0.137& 8.849& 4.0 &  201& 203.55& 1.013 \\
\hline
9 &
$1.9$ & 90.98 & 1.624 &-0.149& 9.696& 4.0 & 201& 202.18 & 1.006 \\
\hline
\end{tabular}
\end{center}
\caption{ BGK fit with valence quarks for $\sigma_r$  for H1ZEUS-NC-(e+p) and H1ZEUS-NC-(e-p) data in the range $Q^2 \ge 3.5$ and $x\le 0.01$. NLO fit. RT HF Scheme. { \it Soft gluon}.}
\label{tabl1}
\end{table}

In this section we investigate how well the dipole model can describe the new, precise, HERA data which were obtained in the region of Q$^2> 3.5$ GeV$^2$.
Since the quality of data in the region of Q$^2 <1$  GeV$^2$ was not improved until now we concentrate here on the higher Q$^2$ region where the valence quark contribution becomes relevant.  
\subsection{Dipole fits with valence quarks}
First, we show  that it is possible to  combine the dipole and valence quark contributions and obtain a good fit to the data. For the purpose of this investigation we choose the BGK model because it uses the DGLAP evolution. The fits were performed within the HERAfitter framework, i.e.;  the QCD evolution is the same as in the standard HERAfitter pdf fits. The results of the BGK fit, with valence quarks, are shown in  Table~\ref{tabl1}. The fit is performed in  the low $x$ range, $x < 0.01$, for various $\mu_0^2$ values.    The value of $\mu_0^2$  plays a role of the starting scale of the QCD evolution which is usually denoted by $Q_0^2$ in the pdf fits. $N_p$ denotes the number of measured values of the reduced cross section, $\sigma_r$, which were used in the fit. The parameters  
 $\sigma_{0}$ of the dipole model and the  parameters for gluon $A_{g}$, $\lambda_{g}$, $C_{g}$ are obtained from the fit at a given value of $Q_0^2$ (in GeV$^2$).  The value of the parameter $C$ was fixed, as explained above. 
%*********************************
%\vskip 1.0cm

The table shows that the BGK model with valence quarks taken from the usual HERAfitter pdf fit, is describing the precise HERA data very well for all $Q_0^2$ value. The fit quality improve slightly with diminishing $Q_0^2$. This could indicate that HERA data in the low range of $Q^2 \sim 3.5$ GeV$^2$, retain some sensitivity to the saturation effects. In the BGK model the saturation effects increases with decreasing $Q_0^2$ value.

In the Table~\ref{tabl2} we show results of the standard HERAPDF fits. They  are performed in the same Q$^2$ range as the dipole fits but in the full $x$ range. The full $x$ range is necessary  to fix the contribution of  valence quarks.

\begin{table}[h]
\begin{center}
\begin{tabular}{|c||c||c||c|c||c|c|c||c|c|c||c|} 
\hline 
No& 
$Q_0^2$&HF Scheme& $\chi^2$&$Np$& $\chi^2/Np$\\
\hline
1 &
1.1& RT & 604.64 & 592& 1.021 \\
\hline
3&
1.3& RT & 586.33 & 592& 0.990 \\
\hline
5&
1.5& RT & 579.72 & 592& 0.979 \\
\hline
7&
1.7& RT & 576.76 & 592& 0.974 \\
\hline
9&
1.9& RT & 575.08 & 592& 0.971 \\
\hline
\end{tabular}
\end{center}
\caption{HERAPDF fit for $\sigma_r$  for H1ZEUS-NC-(e+p), H1ZEUS-NC-(e-p) and H1ZEUS-CC-(e+p), H1ZEUS-CC-(e-p) data in the range $Q^2 \ge 3.5$ and $x\le 1.0$.}
\label{tabl2}
\end{table}
\begin{table}[htbp]
\begin{center}
\begin{tabular}{|c||c||c||c|c||c|c|c||c|c|c||c|} 
\hline 
No& 
$Q_0^2$&HF Scheme& $\chi^2$&$Np$& $\chi^2/Np$\\
\hline
1 &
1.1& RT & 472.52 & 550& 0.859 \\
\hline
3&
1.3& RT & 469.80 & 550&  0.854 \\
\hline
5&
1.5& RT &  469.06 & 550& 0.853 \\
\hline
7&
1.7& RT & 468.67 & 550& 0.852 \\
\hline
9&
1.9& RT & 468.34 & 550& 0.852 \\
\hline
\end{tabular}
\end{center}
\caption{HERAPDF fit for $\sigma_r$  for H1ZEUS-NC-(e+p), H1ZEUS-NC-(e-p) and H1ZEUS-CC-(e+p), H1ZEUS-CC-(e-p) data in the range $Q^2 \ge 8.5$ and $x\le 1.0$.}
\label{tabl3}
\end{table}
 
\begin{table}[htbp]
\begin{center}
\begin{tabular}{|c||c||c||c|c||c|c|c||c|c|c||c|} 
\hline 
No& 
$Q_0^2$&
$\sigma_0$ & $A_g$ & $\lambda_g$ & $C_g$ & $C$ & $Np$& $\chi^2$& $\chi^2/Np$\\
\hline
% h09a & 
1 &
$1.1$ &  91.60  &  2.227 & -0.022 &  9.322 & 4.0 &  162 & 131.78 & 0.813 \\
\hline
3 &
$1.3$ & 83.393 & 2.047 & -0.069& 10.019 & 4.0 &  162 & 132.10 &0.815 \\
\hline
5 &
$1.5$ & 77.121& 1.969 &-0.098& 10.825& 4.0 & 162&  132.23 &  0.816 \\
\hline
7 &
$1.7$ & 71.975& 1.922 &-0.120& 11.538& 4.0 & 162& 132.88& 0.820 \\
\hline
9 &
$1.9$ & 69.128 & 1.897 &-0.135& 12.175& 4.0 & 162&  132.03 & 0.815 \\
\hline
\end{tabular}
\end{center}
\caption{Dipole model BGK fit with valence quarks for $\sigma_r$  for H1ZEUS-NC-(e+p) and H1ZEUS-NC-(e-p) data in the range $Q^2 \ge 8.5$ and $x\le 0.01$. NLO fit. RT HF Scheme. {\it Soft gluon}.}
\label{tabl4}
\end{table}

Table~\ref{tabl2} shows a very good agreement with data of the standard pdf fit. The agreement is similar as in the dipole fits, if corrected  for the number of points and the number of free parameter, which is  $N_{free} =10$ for the HARAPDF fit and  $N_{free} =4$  in case of the BGK fit with the {\it soft} gluon assumption.  In difference to the dipole fits, the quality of the HERAPDF fit is  deteriorating with decreasing $Q_0^2$ scale.   

Table~\ref{tabl3} and~\ref{tabl4}  show HERAPDF and BGK dipole fits in the higher  $Q^2$ range,  $Q^2 > 8.5$ GeV$^2$. We see that  the quality of  fits  clearly improves in the higher $Q^2$ region. In case of HARAPDF fit the $\chi^2/N_p$ improves from  0.97 to 0.85 and in case of the BGK fit from $\sim \, 1.0$ to 0.82.   Moreover,
the BGK fits do not show any dependence from the starting scale, $Q_0^2$. The HERAPDF fits do still show some slight deterioration with decreasing $Q_0^2$ but the effect is much smaller than seen in Table~\ref{tabl2}.   

%The large improvement of the BGK fit indicates that the eikonal approximation used to model saturation effects in the BGK or GBW approach, is too rudimentary.  

\begin{figure}[h]
\centering
\includegraphics[width=9.0cm]{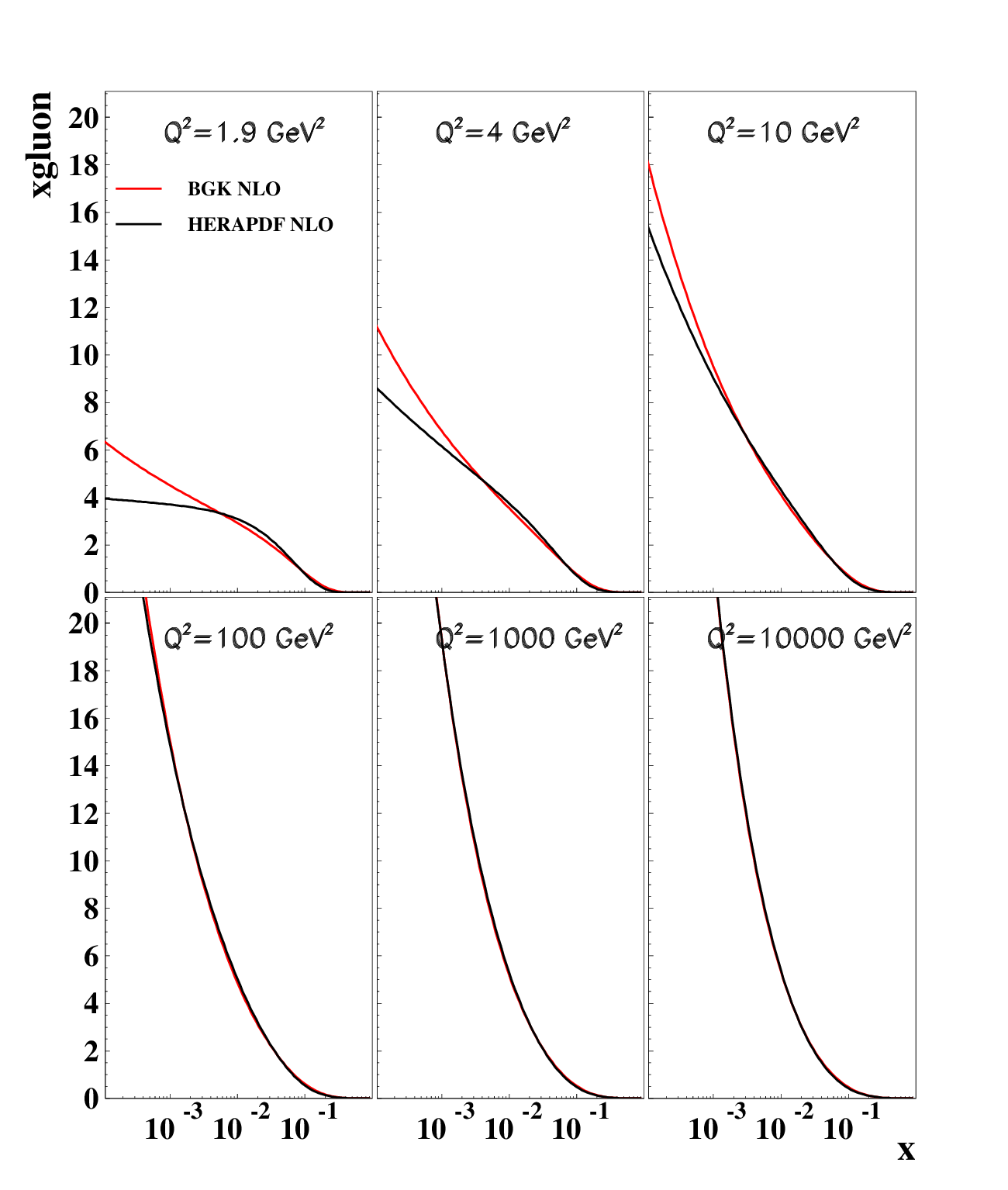}
\caption{Comparison between the dipole (soft)  and HERAPDF gluon in NLO. }
\label{glu-nlo}
\end{figure}

\begin{figure}[h]
\centering
\includegraphics[width=9.0cm]{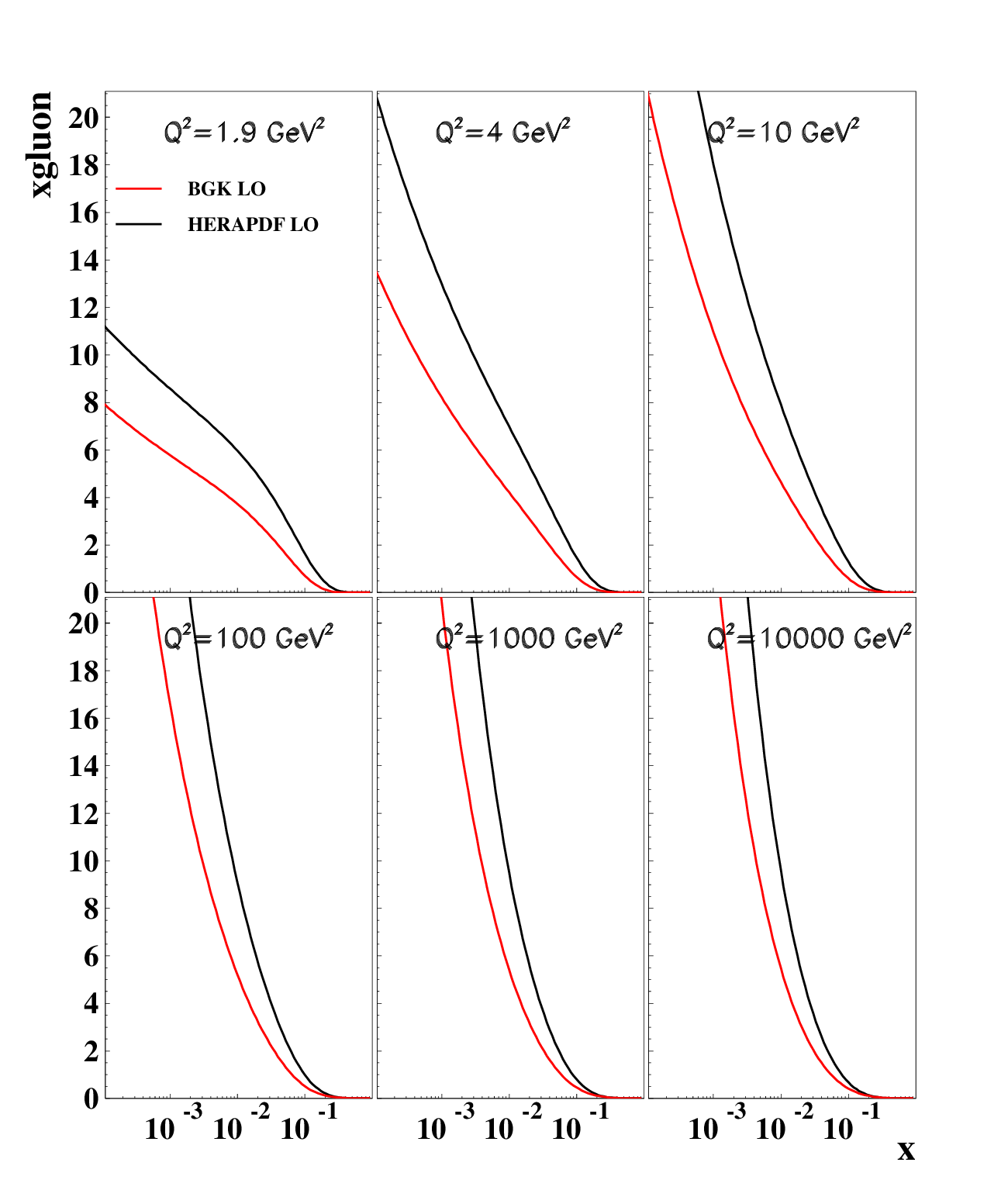}
\caption{Comparison between the dipole (soft)  and HERAPDF gluon in LO. }
\label{glu-lo}
\end{figure}

In Fig.~\ref{glu-nlo} we show a comparison the gluon density obtained in the fits with valence quarks and compare it to the gluon density obtained in the HERAPDF fit.  We see that the two gluon densities, at NLO, differ at smaller scales but then start to approach each other at higher scales. It is interesting to observe that the convergence of the two gluon densities is much slower in LO, Fig.~\ref{glu-lo}.    

\subsection{Fits with  alternative forms of the gluon density}
In this section we investigate whether the more involved  forms of the  gluon density,  eq.(\ref{gden-softhard}) and eq.(\ref{gden-softneg}), can improve  the data description.  In table \ref{tabl5} and \ref{tabl6} we show the fit results for the fits  with  {\it soft + hard}  gluon of eq.(\ref{gden-softhard}), in the lower $Q^2 > 3.5$ GeV$^2$ and higher $Q^2 > 8.5$ GeV$^2$ regions. We observe that the fit quality improves significantly by adding a "hard" component, $D_gx +E_gx$, to a classic  soft gluon of eq.(\ref{gden-soft}). The value of $\chi^2$ diminishes by about $\Delta \chi^2 \approx 20$ for $Q^2 >3.5$ GeV$^2$ and by about  $\Delta \chi^2 = 15$ for $Q^2>8.5$ GeV$^2$, which is a much larger drop than the increase of the  parameter number (just  by 2). 

\begin{table}[htbp]
\begin{center}
\begin{tabular}{|c||c||c||c|c||c|c|c||c|c|c||c|} 
\hline 
No& 
$Q_0^2$&
$\sigma_0$ & $A_g$ & $\lambda_g$ & $C_g$& $D_g$& $E_g$ & $\chi^2$& $\chi^2/Np$\\
\hline
% h09a & 
1 &
$1.1$ &  217.09  & 1.976 & -0.012 &  22.502 &-35.364 & 1339.3&  181.34 & 0.930 \\
\hline
2 &
$1.3$  & 181.82  & 1.847 & -0.059 &  21.597 &-25.051 &1030.3 &  180.80 & 0.927\\
\hline
3 &
$1.5$ & 165.17  & 1.871 & -0.082 &  24.623& -23.630 &1237.7 & 180.80 & 0.927 \\
\hline
4 &
$1.7$ & 147.12  & 1.903 & -0.099 &  26.720 &-20.584 &1310.2 &  181.70 & 0.932 \\
\hline
5 &
$1.9$ & 132.26 & 1.948 &-0.111& 28.211 &-18.008 &1322.4  & 180.81 & 0.927 \\
\hline
\end{tabular}
\end{center}
\caption{Dipole model BGK fit with valence quarks for $\sigma_r$  for H1ZEUS-NC-(e+p) and H1ZEUS-NC-(e-p) data in the range $Q^2 \ge 3.5$ and $x\le 0.01$. NLO fit. RT HF Scheme. {\it Soft + hard gluon}.  $Np =201$ and C=4.0 GeV$^2$. }
\label{tabl5} 
\end{table}

\begin{table}[htbp]
\begin{center}
\begin{tabular}{|c||c||c||c|c||c|c|c||c|c|c||c|} 
\hline 
No& 
$Q_0^2$&
$\sigma_0$ & $A_g$ & $\lambda_g$ & $C_g$ & $D_g$& $E_g$ & $\chi^2$& $\chi^2/Np$\\
\hline
% h09a & 
1 &
$1.1$ &  254.97  & 2.524 & -0.027 &  24.857 &-46.523 & 1639.8& 117.34 & 0.752 \\
\hline
2 &
$1.3$ & 154.25 & 2.171 & -0.041 & 13.728 & -20.261 &340.97 &  121.79 & 0.781 \\
\hline
3 &
$1.5$ & 292.89  & 2.358 & -0.034 &31.168 & -50.312& 2585.8 &  115.51 & 0.740 \\
\hline
4 &
$1.7$ &221.52 & 2.483 &-0.051& 34.010&-44.156 &2630.6 & 115.78 & 0.742 \\
\hline
5 &
$1.9$ & 174.46 & 2.490 &-0.070& 35.347 &-37.706& 2499.7& 116.18 & 0.745 \\
\hline
\end{tabular}
\end{center}
\caption{Dipole model BGK fit with valence quarks for $\sigma_r$  for H1ZEUS-NC-(e+p) and H1ZEUS-NC-(e-p) data in the range $Q^2 \ge 8.5$ and $x\le 0.01$. NLO fit. RT HF Scheme. {\it{Soft + hard gluon}}.  $Np =162$ and $C=4.0$ GeV$^2$. }
\label{tabl6}
\end{table}

In Table~\ref{tabl7} we show  the fit results for the fits with the {\it soft + negative} gluon of eq.({\ref{gden-softneg}). The fit in the lower $Q^2$ range is not significantly improved by the addition of the negative gluon term. In the higher $Q^2$ range, $Q^2> 8.5$ GeV$^2$, the fit improves somewhat, although not so clearly as in the "hard" case.

\begin{table}[htbp]
\begin{center}
\begin{tabular}{|c||c||c||c|c||c|c|c||c|c|c||c|} 
\hline 
$Q_0^2$&$Q^2$ &
$\sigma_0$ & $A_g$ & $\lambda_g$ & $C_g$& $A'_g$& $B'_g$&$C'_g$ & $\chi^2$& $\chi^2/Np$\\
\hline
% h09a & 
$1.9$ & $3.5$ &115.09 & 0.874 & -0.253 & 3.669&-0.014& -0.606& 25.0& 200.49 & 1.028 \\
\hline
$1.9$  &$8.5$ &111.94  & 0.799 & -0.290 &  3.922 &0.020 &-0.642& 25.0 &  119.48 & 0.766\\
\hline
\end{tabular}
\end{center}
\caption{Dipole model BGK fit with valence quarks for $\sigma_r$  for H1ZEUS-NC-(e+p) and H1ZEUS-NC-(e-p) data. NLO fit. RT HF Scheme. {\it Soft + negative gluon}.  $C =4.0$ GeV$^2$, $N_p =201$ for $Q^2>3.5$ GeV$^2$ and $N_p =162$ for $Q^2>8.5$ GeV$^2$}
\label{tabl7}
\end{table}

\subsection{Fits without or with fitted valence quarks}
To better understand the meaning of the fits which are using alternative forms of the gluon density   we performed also fits without valence quarks, and with valence quarks fitted to data.  In Table~\ref{tabl8} and Table~\ref{tabl9} we show fits performed without valence quarks for the soft and soft+hard forms of the gluon density in the region of $Q^2 > 3.5$ GeV$^2$.
\begin{table}[ht]
\begin{center}
\begin{tabular}{|c||c||c||c|c||c|c|c||c|c|c||c|} 
\hline 
No& 
$Q_0^2$&$Q^2$ &
$\sigma_0$ & $A_g$ & $\lambda_g$ & $C_g$  & $\chi^2$& $\chi^2/Np$\\
\hline
1 &
$1.9$ & $3.5$  &115.09  & 2.038 & -0.097 & 4.969&  197.83 & 1.004 \\
\hline
\end{tabular}
\end{center}
\caption{Dipole model BGK fit without valence quarks for $\sigma_r$  for H1ZEUS-NC-(e+p) and H1ZEUS-NC-(e-p) data in the range $Q^2 \ge 3.5$ and $x\le 0.01$. NLO fit. RT HF Scheme. {\it{Soft gluon}}.  $C =4.0$ GeV$^2$ and $Np =201$. }
\label{tabl8}
\end{table}
\begin{table}[htbp]
\begin{center}
\begin{tabular}{|c||c||c||c|c||c|c|c||c|c|c||c|} 
\hline 
No& 
$Q_0^2$&$Q^2$&
$\sigma_0$ & $A_g$ & $\lambda_g$ & $C_g$ & $D_g$&$E_g$ & $\chi^2$& $\chi^2/Np$\\
\hline
1 &
$1.9$ & $3.5$& 119.18  & 1.970 & -0.104 & 5.001 &3.347&-19.340 & 196.26 & 1.006 \\
\hline
\end{tabular}
\end{center}
\caption{Dipole model BGK fit without valence quarks for $\sigma_r$  for H1ZEUS-NC-(e+p) and H1ZEUS-NC-(e-p) data in the range $Q^2 \ge 3.5$ and $x\le 0.01$. NLO fit. RT HF Scheme. {\it {Soft + hard gluon}}.  $C =4.0$ GeV$^2$ and $Np =201$. }
\label{tabl9}
\end{table}

The contribution of the valence quarks in the low $x$ region are large enough to be able to determine them in this region only. In Table~\ref{tabl10} we show an example of a fit with  parameters of the valence quarks fitted to data together with the parameters of the gluon density. The fit is performed for $Q^2 >3.5$ GeV$^2$, in the low $x$ range, $x<0.01$.  In Table~\ref{tabl11} we give, for completeness, the parameters of the valence quarks determined in this way.  Note, that the fit with fitted valence quarks is better than the fit with fixed valence quarks of Table~\ref{tabl1} and it is also better than the fit without valence quarks, Table~\ref{tabl8}.

Fig.~\ref{glu-softhard} shows the comparison between the NLO gluon densities determined with the {\it soft} and {\it soft + hard assumptions}. 
 The  {\it soft} gluon density is taken from the fit  of  Table~\ref{tabl1}.  The  {\it soft+hard} gluon density shown on the LHS of Fig.~\ref{glu-softhard}  is taken from the fit  of  Table~\ref{tabl5} and was obtained with the fixed valence quark contribution. The RHS of this figure shows the  {\it soft+hard} gluon density obtained from the fit of   Table~\ref{tabl10}. Here, the  contribution of valence quarks is fitted to data together with the gluon density. Both fits, of   Table~\ref{tabl5} and Table~\ref{tabl10}, have a very similar quality, the form of gluon densities differs, however, at lower scales in the high $x$ region; the one with the fixed valence quarks shows a clear bump around $x\approx 0.1$, the another  shows no bump and has a form similar to the {\it soft} case. In all fits which we performed, the bump in the {\it soft+hard} gluon density fitted with the fixed valence quarks  was always present.  independently of the $Q^2$ cut  or the LO or NLO  QCD evolution. This bump disappears, however, when the valence quark contribution is fitted. Therefore, we do not attribute a physical meaning to this bump, especially that it is in the region which is not directly tested by data and it contributes only to the low-$x$ region through the QCD evolution.    Nevertheless, its existence emphasizes the necessity of a full fit to the data, i.e. of a fit in which the gluon density is fitted  {\it together} with the valence quarks. 
\begin{figure}[htbp]
\centering
\includegraphics[width=8.0cm]{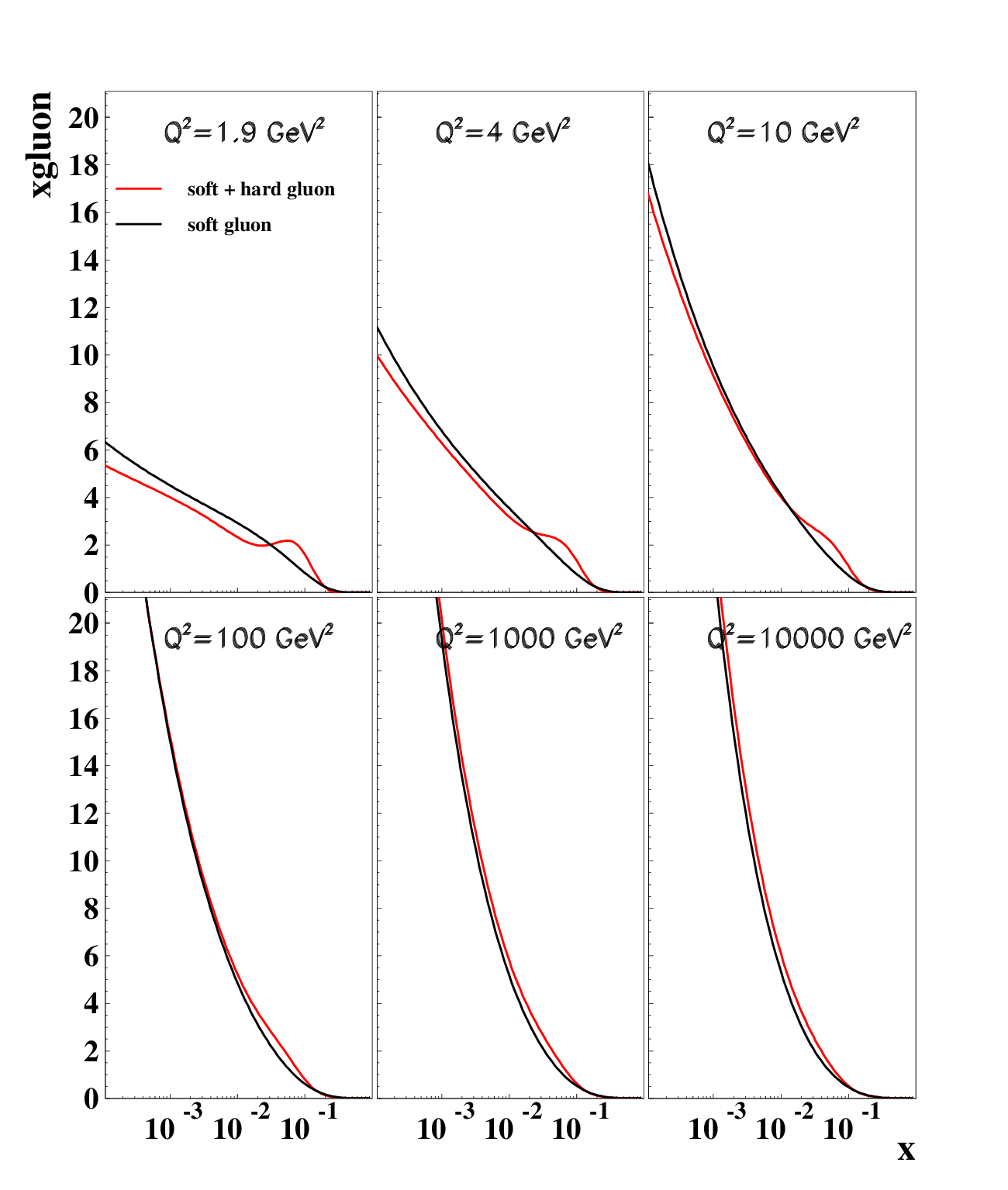}
\includegraphics[width=8.0cm]{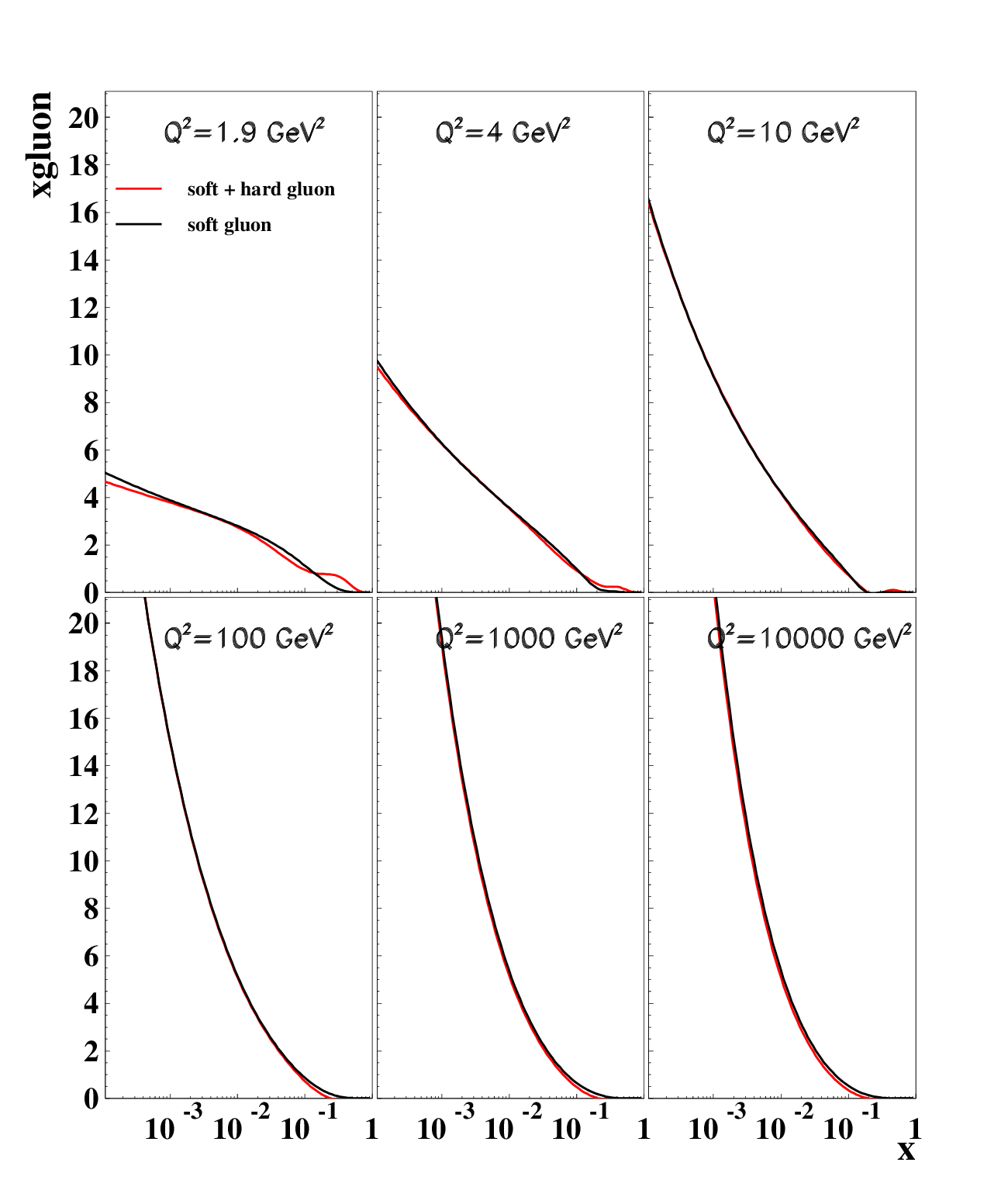}
\caption{Comparison between the NLO gluon densities determined with the soft and soft + hard assumptions. LHS shows the gluon distribution functions determined with the fixed valence quark contribution. RHS shows the gluon distribution functions determined with the contribution of valence quarks fitted to data in the $x<0.01$ region.   }
\label{glu-softhard}
\end{figure}

\begin{table}[htbp]
\begin{center}
\begin{tabular}{|c||c||c||c|c||c|c|c||c|c|c||c|} 
\hline 
No& 
$Q_0^2$&$Q^2$ &
$\sigma_0$ & $A_g$ & $\lambda_g$ & $C_g$  & $\chi^2$& $\chi^2/Np$\\
\hline
1 &
$1.9$ & $3.5$  &88.040  & 1.766 & -0.115 & 6.747&  182.89 & 0.978 \\
\hline
\end{tabular}
\end{center}
\caption{Dipole model BGK fit with valence quarks fitted for $\sigma_r$  for H1ZEUS-NC-(e+p) and H1ZEUS-NC-(e-p) data in the range $Q^2 \ge 3.5$ and $x\le 0.01$. NLO fit. RT HF Scheme. {\it {Soft gluon}}.  $C =4.0$ GeV$^2$ and $Np =201$. }
\label{tabl10}
\end{table}
\begin{table}[htbp]
\begin{center}
\begin{tabular}{|c||c||c||c|c||c|c|c||c|c|c||c|} 
\hline 
No& 
$ Auv$& $Buv$ & $Cuv$&  $Euv$ & $Adv$&
$Cdv$ & $CUbar$ & $ADbar$ & $BDbar$& $CDbar$\\
\hline
1 &
3.717& 0.665 & 4.652 &9.694& 2.189 & 4.291 & 2.582&0.100& -0.165 & 2.405\\
\hline
\end{tabular}
\end{center}
\caption{Parameters for valence quarks from : dipole model BGK fit with valence quarks fitted for $\sigma_r$  for H1ZEUS-NC-(e+p) and H1ZEUS-NC-(e-p) data in the range $Q^2 \ge 3.5$ and $x\le 0.01$. NLO fit. RT HF Scheme. {\it {Soft gluon}}. Parameter $cBGK =4.0$, $Np =201$. }
\label{tabl11}
\end{table}

\section{Comparison with HERA data}
In Fig.~\ref{bgk-heraf2} we show a comparison of the dipole BGK fit  with the  HERA reduced cross section data. Figure shows an excellent agreement of the fit with data. In Fig.~\ref{bgk-herafl} we show  a comparison of $F_l$ structure function obtained from the dipole BGK fit  with   HERA data. In both figures we use the BGK fit of  Table~\ref{tabl1}, with $Q_0^2=1.9$ GeV$^2$. Very similar comparisons but calculated from different fits with $\mu_0^2$ value fitted, were done in the conference proceedings \cite{Luszczak:2013vta}, being a  short introduction to the studies presented here.  
\begin{figure}[htbp]
\centering
\includegraphics[width=8.1cm]{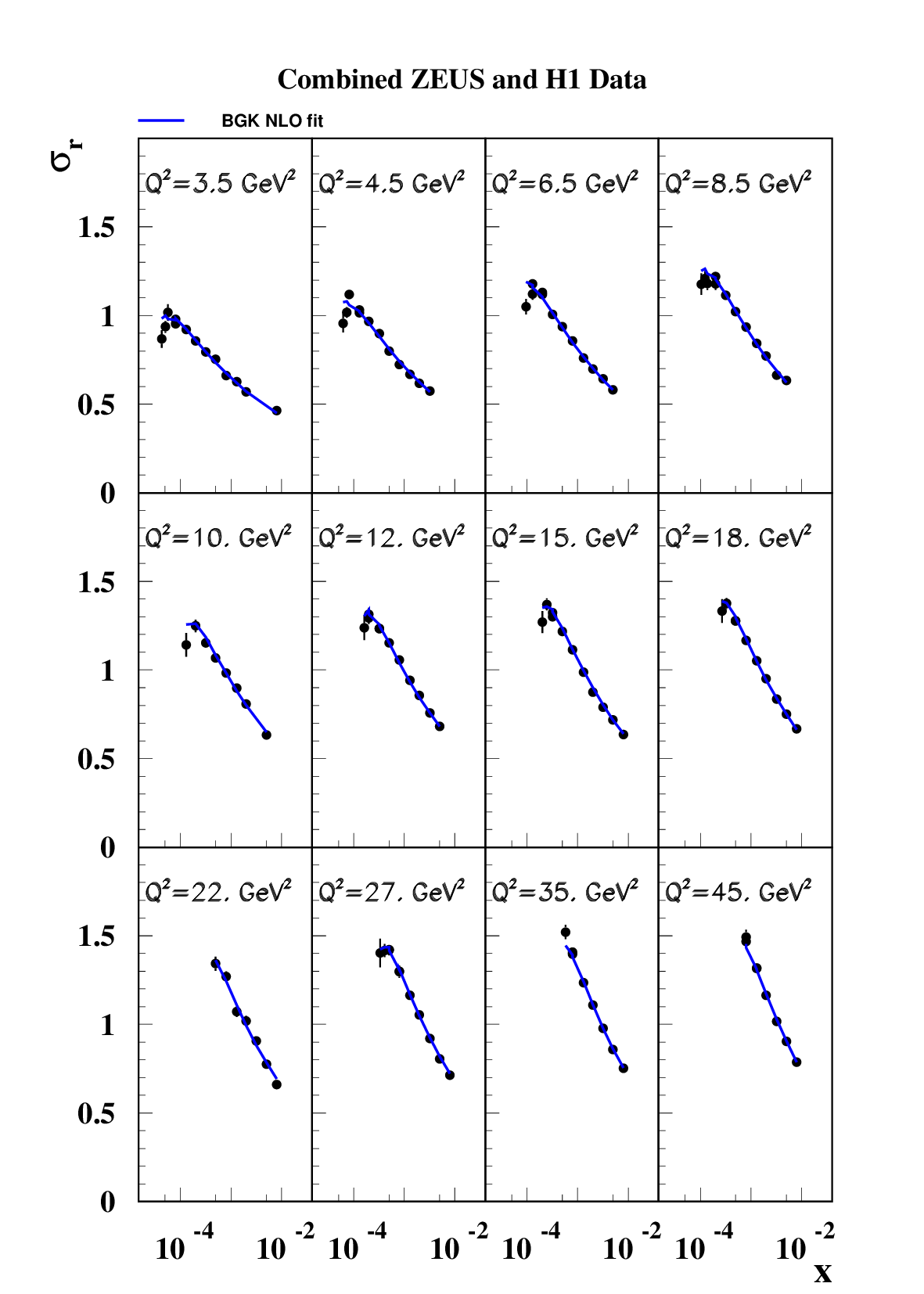}
\includegraphics[width=8.1cm]{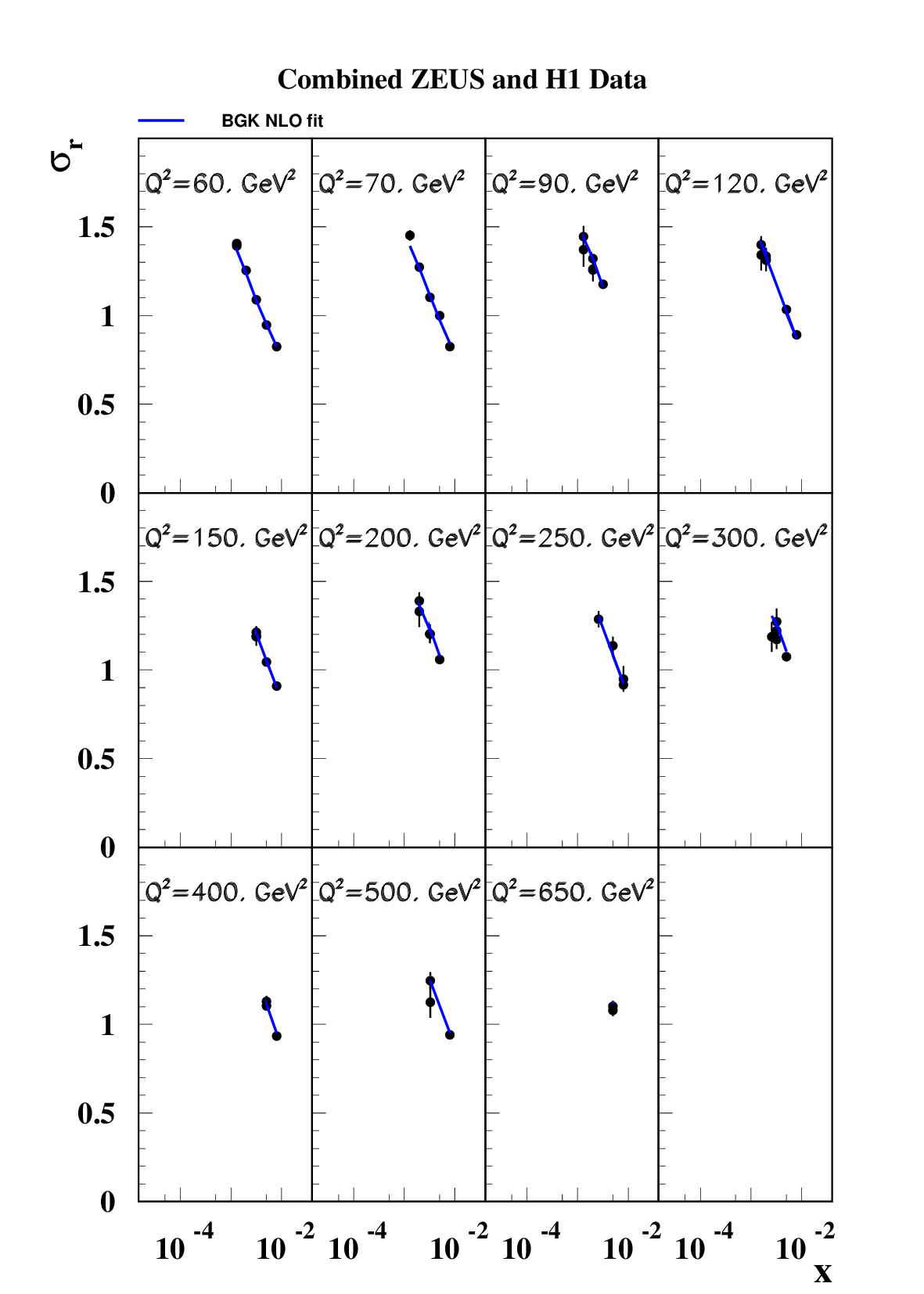}
\caption{Comparison  of the dipole BGK fit of Table~\ref{tabl1} with the reduced cross sections  of  HERA data.}
\label{bgk-heraf2}
\end{figure}
\begin{figure}[htbp]
\centering
\includegraphics[width=12.0cm] {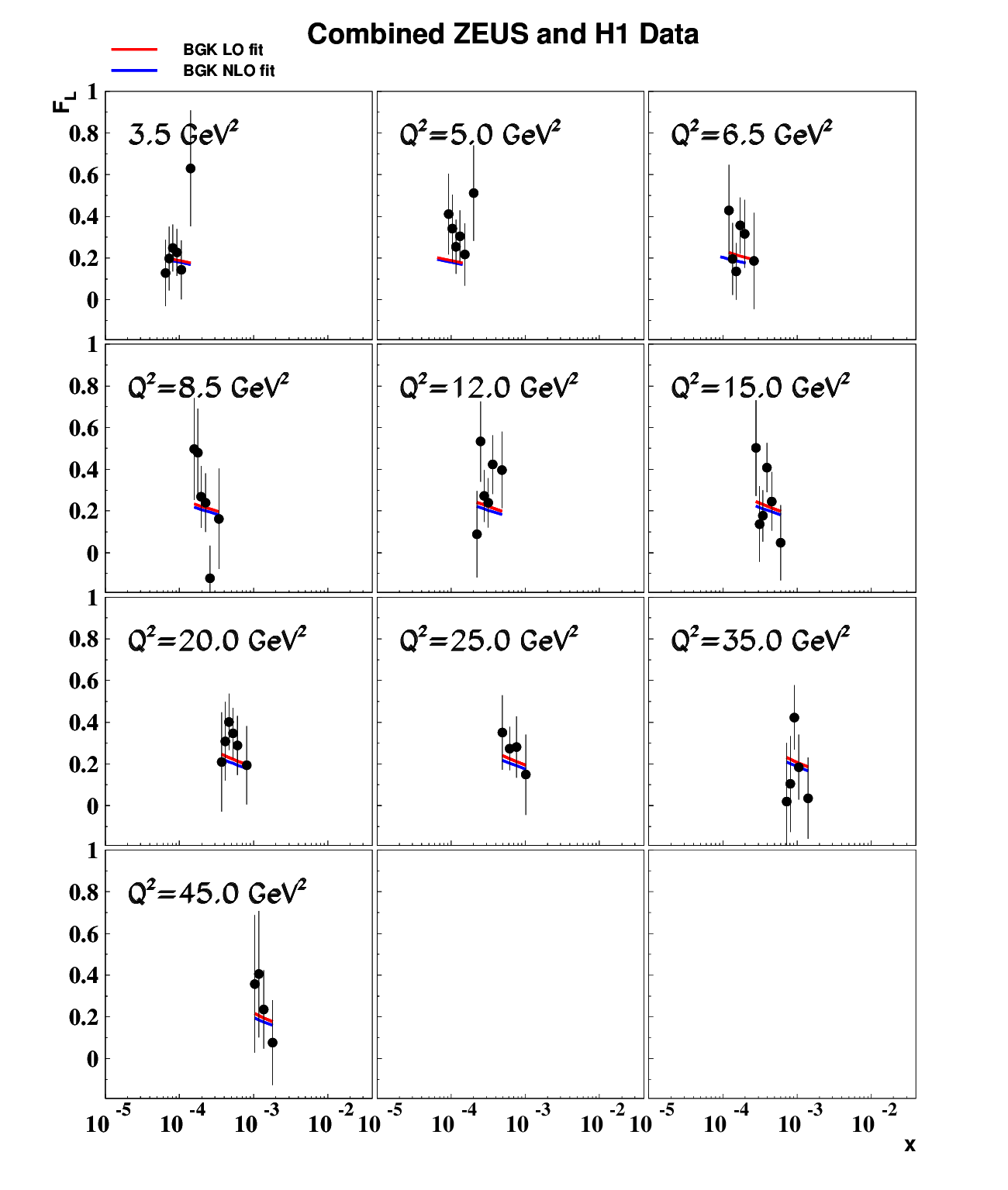}
\caption{Comparison  of  $F_l$ structure function obtained from the dipole BGK fit of Table~\ref{tabl1} with  HERA data.  }
\label{bgk-herafl}
\end{figure}

\newpage

\section{Summary}

We have shown that the $k_T$ factorized, DGLAP evolved gluon density, evaluated within the BGK model, describe the combined, precise HERA data in the low-$x$ region, very well.
The valence quark contribution added to the dipole model improves the fit  significantly. Therefore, for precise dipole evaluations the gluon contribution should be complemented by valence quarks.

The resulting gluon density obtained from fits with fitted valence quarks could be  used for the prediction of LHC cross sections, provided that the dipole amplitude, which is now only well defined in the low-$x$ region, can be analytically continued to the high $x$ region~\cite{KLRn}. 

As a byproduct of this investigation we observe that
the fits of all dipole and pdf types improve  significantly when the $Q^2$ cut on data is increased from $Q^2>3.5$ to  $Q^2>8.5$ GeV$^2$. We have checked this with the dipole model with quarks and without quarks, with various forms of the gluon density, as well as with the standard HERAPDF1.0 fit.
The persistence of this effect indicate some shortcomings of the theoretical description; it could be due to the lack of higher order QCD corrections or to  saturation effects. We note,  that the higher order corrections diminish logarithmically with increasing $Q^2$ whereas the saturation effects diminish  like a power of $Q^2$~\footnote{The degree of saturation is characterized by the size of the dipole, $r_S$, which, at a given $x$,  starts to interact multiple times (in about 60\% of cases). 
We recall that the saturation scale, $Q_S^2=2/r_S^2$,  was determined at HERA as 0.5 GeV$^2$ at $x=10^{-3}$ and as about 1 GeV$^2$ at $x=10^{-4}$~\cite{KMW}.  Therefore, to avoid  multiple scattering of dipoles,  the $Q^2$ cut should be by about a factor of 10 higher than the saturation scale. 
}, or  faster.  In our view,  the relatively fast  change of  $\chi^2/N_p$ with the increased $Q^2$ cut indicates that the effect is due to saturation, at least to large extent. In this way, the increase of precision in HERA data offers a novel testing ground for saturation study in the well measured region above $Q^2>3.5$  GeV$^2$.  The study of this type may become very interesting when, in the near future, the combined HERA I and HERA II data,  with yet further increased precision, is published.

% which are apparently not well described by the eikonal approximation used in the BGK dipole model.
%This could suggest that  a more involved saturation mechanism should be considered, like in the IIM model~\cite{Iancu:2003ge}. The prerequisite of application of such a model at higher $Q^2$ is, however, a precise transition to DGLAP in this region. 

\section{Acknowledgement}
We would like to thank P. Belov,  A. Glazov, R. Plackacyle and Voica Radescu for introduction to the HERAfitter project and various help with solving  problems.  We would  also  like to thank   J. Bartels and D. Ross for reading the manuscript and useful comments.  
%\end{acknowledgments}

\end{document}